\documentclass[conference]{IEEEtran}

\usepackage{cite}
\usepackage{graphicx}
\usepackage[cmex10]{amsmath}
\usepackage{array}
\usepackage[font=footnotesize,caption=false]{subfig}
\usepackage[hyphens]{url}
\usepackage{cprotect}
\usepackage{comment}
\usepackage{color}
\usepackage{amsfonts,amssymb,amsmath}
\usepackage{tikz}
\usetikzlibrary{calc}
\usepackage{booktabs}
\usepackage{footmisc}
\usepackage{hyperref}
\hypersetup{colorlinks=true,linkcolor=blue}

\newcommand{\Z}{\mathbb{Z}}
\newcommand{\wrt}{w.\thinspace r.\thinspace t.\ }
\DeclareMathOperator{\rmax}{rmax}

\begin{document}

\title{Pentago is a First Player Win: Strongly Solving a Game Using Parallel In-Core Retrograde Analysis}
\author{\IEEEauthorblockN{Geoffrey Irving}\IEEEauthorblockA{Otherlab\\San Francisco, CA\\irving@naml.us}}

\maketitle

\begin{abstract}
\boldmath
We present a strong solution of the board game pentago, computed using exhaustive parallel
retrograde analysis in 4 hours on 98304 ($3 \times 2^{15}$) threads of NERSC's Cray Edison.
At $3.0 \times 10^{15}$ states, pentago is the largest divergent game solved to date by
two orders of magnitude, and the only example of a nontrivial divergent game solved
using retrograde analysis.
Unlike previous retrograde analyses, our computation was performed entirely in-core,
writing only a small portion of the results to disk; an out-of-core implementation would
have been much slower.  Symmetry was used to reduce branching factor
and exploit instruction level parallelism.  Despite a theoretically
embarrassingly parallel structure, asynchronous message passing was required to fit the
computation into available RAM, causing latency problems
on an older Cray machine.  All code and data for the project are
open source, together with a website which combines database lookup and on-the-fly computation
to interactively explore the strong solution.
\end{abstract}

\IEEEpeerreviewmaketitle

\vspace{-.02in}
\section{Introduction}

Computer play of combinatorial games such as chess, checkers, and go has been an active area of research
since the early days of computer science \cite{shannon1950chess}.  The limit of computer play is a solved
game, when a computer can play perfectly either from the start position (weakly solved) or from any
position (strongly solved).  The first nontrivial weakly solved game was Connect-Four in 1988 by both
Allen and Allis \cite{allis1988connectfour}, later strongly solved by Tromp \cite{tromp1995}.  Many games
have been solved since, the most challenging being the weak solution of checkers \cite{schaeffer2007checkers}.
The checkers solution involved 18 years of parallel out-of-core retrograde analysis culminating in a
$3.9 \times 10^{13}$ position endgame database together with a $10^{14}$ operation forward search.

To date, all solved games have been either convergent (fewer positions near the end of the game) or amenable
to knowledge-based strategies.  Checkers is an example of a convergent game: while the
entire $10^{20}$ state space is too large to explore fully, the set of positions with 10 or fewer pieces
has a more manageable $3.9 \times 10^{13}$ positions.  Pieces are removed but never added, so a database
of $\le 10$ piece positions can be computed via retrograde (backward) analysis starting with 1 piece, then 2 pieces,
and so on up to 10 pieces.  The computed database is then used to prune a forward search starting from the
beginning of the game.

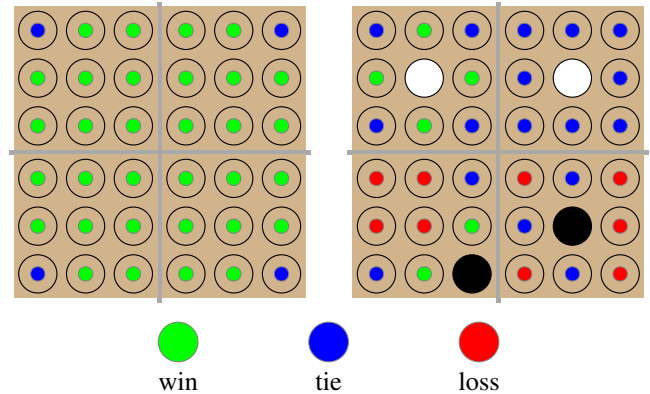
\begin{figure}
\begin{center}
\includegraphics[width=.45\columnwidth]{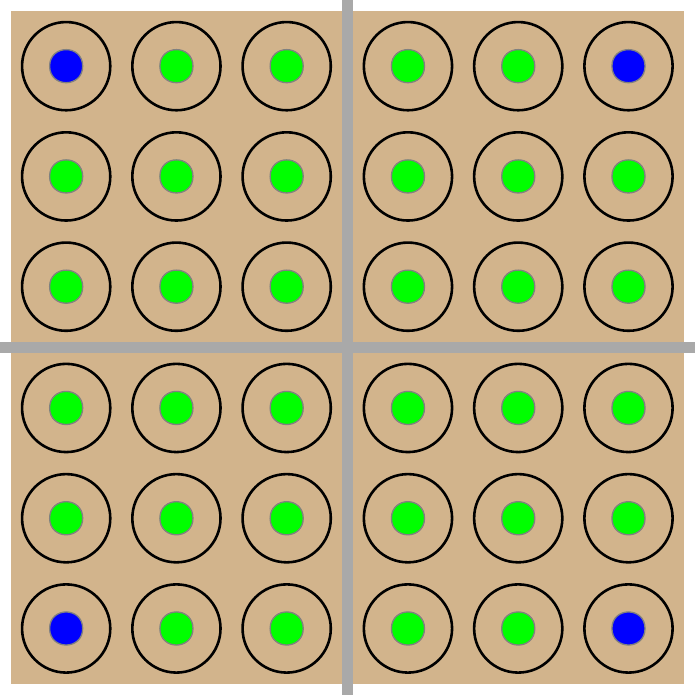}
\hspace{.1in}
\includegraphics[width=.45\columnwidth]{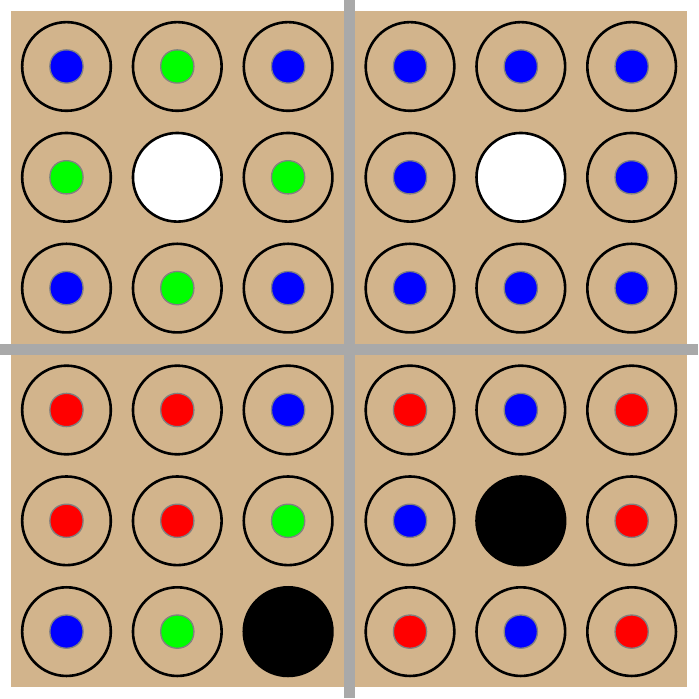} \\
\vspace{.1in}
\begin{tikzpicture}
  \definecolor{win} {rgb}{0,1,0};
  \definecolor{tie} {rgb}{0,0,1};
  \definecolor{loss}{rgb}{1,0,0};
  \tikzstyle{kind}=[circle,draw=gray,inner sep=0pt,minimum size=15]
  \node at (-2,0) [kind,label=below:win, fill=win] {};
  \node at ( 0,0) [kind,label=below:tie, fill=tie] {};
  \node at ( 2,0) [kind,label=below:loss,fill=loss] {};
\end{tikzpicture}
\end{center}
\vspace{-.14in}
\cprotect\caption{(Left) With perfect play, the first player wins with any opening move except the
corners, which tie.  (Right) A more delicate position with black to play.
The full strong solution can be explored at \url{http://perfect-pentago.net}.}
\label{opening}
\vspace{-.04in}
\end{figure}

\begin{figure}
\begin{center}
\includegraphics[width=\columnwidth]{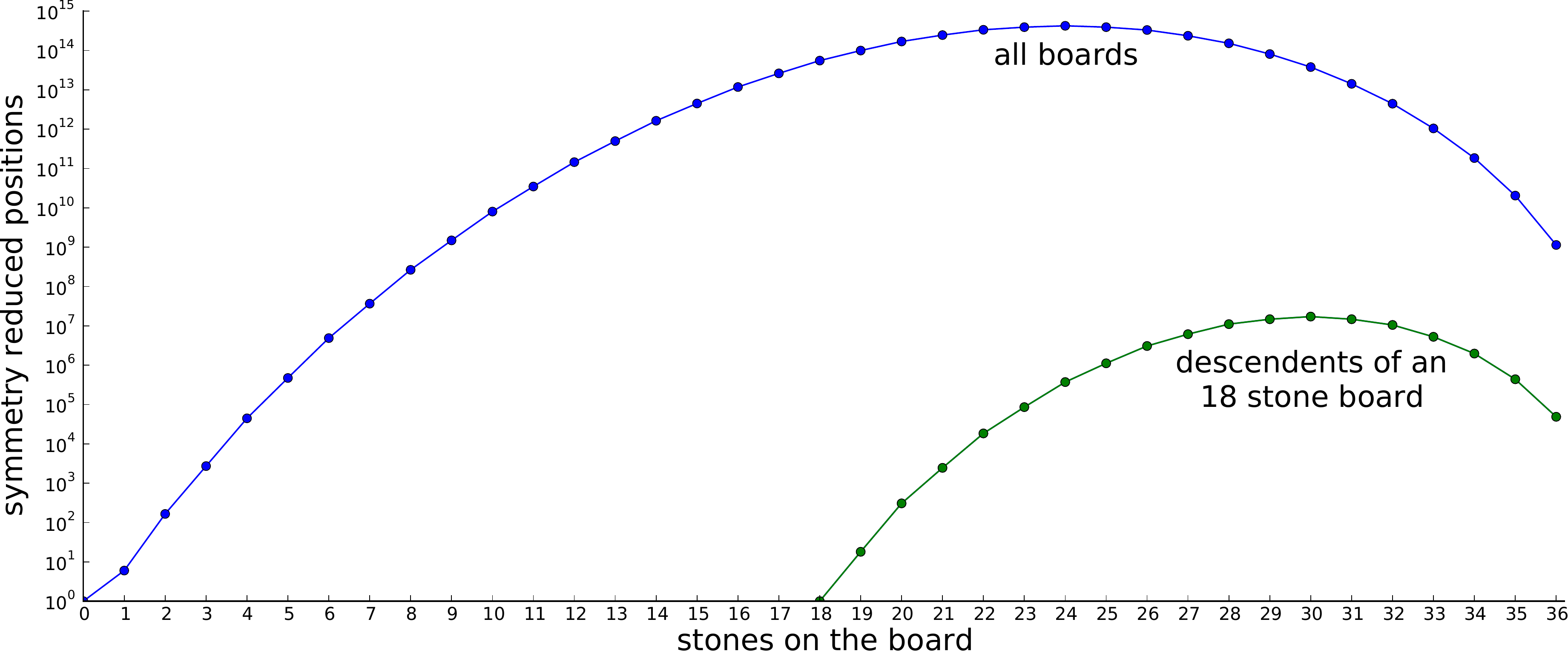}
\end{center}
\vspace{-.1in}
\cprotect\caption{Counts of pentago positions vs.\ stones on the board, with symmetries removed.
Run \verb+web/counts+ in the source repository to reproduce.}
\label{counts}
\vspace{-.1in}
\end{figure}

In contrast to convergent games, the number of positions in a divergent game increases with time
(typically as more stones are added to the board), making traditional retrograde analysis plus forward
search impractical.  Thus, all nontrivial
divergent games solved to date have game-specific knowledge based strategies which can be used to avoid brute
force: zuzgwang control in Connect-Four \cite{allis1988connectfour}, threat-space search for gomoku and renju
\cite{allis1993gomoku,wagner2001renju}, and H-search in hex \cite{arneson2011hex}.  For discussion
on the characteristic of various solved games, see \cite{van2002games}.

Pentago is a divergent game designed by Tomas Flod\'en and sold by Mindtwister \cite{mindtwister2013pentago}.
We reproduce the rules here for completeness.  Pentago is played on a $6 \times 6$ board, divided into four
$3 \times 3$ quadrants. There are two players, black and white, who alternate turns. The goal of each
player is to get five stones of their color in a row, either horizontally, vertically, or diagonally.
Each turn, a player places a stone in an empty space in some quadrant, then chooses a possibly different
quadrant to rotate 90 degrees left or right. If both players get five in a row at the same time, or the
last move is played with no five in a row, the game is a tie. If a player makes five a row by placing a
stone, there is no need to rotate a quadrant: the player wins immediately.

Unlike divergent games solved to date, no strong knowledge based strategies are known for pentago, and
existing programs are capable of searching only to fairly low depths \cite{buescher2009pentago,ewalds2012havannah}.
This is primarily a consequence of the high branching factor of pentago: there are
$36 \cdot 8 = 288$ possible first moves including rotation and an average branching factor of $97.3$ over all
states.\cprotect\footnote{To reproduce the branching factor average, run \verb+bin/analyze branch+ in
\url{https://github.com/girving/pentago}.  For the rest of the paper, only the command will be given.}
To reduce the branching factor to a manageable level, our solver performs all computations in terms of
\emph{rotation abstracted positions} consisting of all 256 ways to rotate the quadrants of a given board;
this eliminates the factor of 8 due to rotation for an average branching factor of only $12.2$.  Operating
on more than one board at a time lets us take full advantage of SSE acceleration.

Unfortunately, symmetry techniques alone are insufficient to solve pentago on commodity hardware.  The game has
3,009,081,623,421,558 ($3 \times 10^{15}$) states with symmetries removed, all but $0.3\%$ of which are reachable
with valid play;\cprotect\footnote{Run \verb+analyze counts+ and \verb+analyze reachable+, respectively.} the number of
states over time is shown in \autoref{counts}.  To solve the game using retrograde analysis, we traverse all
positions in reverse order of the number of stones, starting from the 35 stone \emph{slice} (the 36th is computed
on demand) and iteratively computing the $n$-stone slice from the $(n+1)$-stone slice up to the beginning of the game.
This requires storing two adjacent slices at a time, requiring $213$ TB at peak before compression.  Our initial
target was to fit into half of the NERSC Cray Hopper's $217$ TB, which was plausible using fast but weak compression
only if minimal memory was wasted communication buffers and working storage.

In order to minimize working memory, our parallel solver grabs inputs from other processes immediately before
they are used, overlapping a small number of work chunks to hide latency.  Since computing each chunk takes a variable
amount of time (see below), we opted for a fully asynchronous communication
pattern: when a process needs an input block, it sends a message to the owner of that block, and the owner
replies asynchronously with the data.

The solver was run exactly once at full scale, generating
a $3.7$ TB database of perfect results with $0$ through $18$ stones and establishing that pentago is a win for
the first player to move (\autoref{opening}).  The full strong solution can be explored online at
\url{http://perfect-pentago.net}.

At $3.0 \times 10^{15}$ states, pentago is the largest divergent game computation by a factor of
150 (vs.\ $2 \times 10^{13}$ for $9 \times 6$ Connect-Four), and the largest strongly solved game by a factor of
660 (vs.\ $4.5 \times 10^{12}$ for $7 \times 6$ Connect-Four).  Among retrograde analyses used to solve games,
it is the largest by state space by a factor of 77 (vs.\ $3.9 \times 10^{13}$ in the solution of Checkers).
However, it is not the largest endgame database over any game: the 7-piece Lomonosov Endgame Tablebases for
chess are 140 TB in size, and were computed over six months at Moscow State's Lomonosov supercomputer
\cite{makhnychev2012chess}.  Unfortunately, the technical details of the Lomonosov computation are unpublished,
so a detailed comparison is difficult.

\section{Problem definition}

Let $S$ be the set of arrangements of black and white stones on a $6 \times 6$ board.  Only some of these are
valid pentago positions: if we let black play first, we have equal numbers of black and white stones on black's
turn and one extra black stone on white's turn.  Define predicates $f_b,f_w : S \to \{0,1\}$ by $f_c(s) = 1$
if color $c$ has a five in a row.  Given color $c$, let $\bar{c}$ be the other color.  For $s \in S$, let
$p_c(s) \subset S$ be the positions reached by placing a stone of color $c$, $r(s)$ the positions reached by
rotating exactly one quadrant $90^\circ$ left or right.  Let $v_c(s)$ be the value of position $s$ with $c$
to play: $v_c(s) = -1,0,1$ if $c$ loses, ties, or wins, respectively.  If $f_b(s)$, $f_w(s)$, or $s$ has 36 stones,
the game is over and $v_c(s) = f_c(s) - f_{\bar{c}}(s)$.  Otherwise
\begin{align*}
v_c(s) = \max_{a \in p_c(s)} \begin{cases}
  1 & \mbox{if } f_c(a) \\
  \max_{b \in r(a)} h_c(b) & \mbox{otherwise}
\end{cases}
\end{align*}
where
\begin{align*}
h_c(s) &= \begin{cases} f_c(s) - f_{\bar{c}}(s) & \mbox{if } f_c(s) \wedge f_{\bar{c}}(s) \\
                        -v_{\bar{c}}(s) & \mbox{otherwise} \end{cases}
\end{align*}

\begin{figure*}[ht!]
\begin{center}
\begin{tikzpicture}[scale=.7]
\usetikzlibrary[decorations.pathreplacing]
\draw [decoration={brace,amplitude=7},decorate] (-13.5,4) -- node [above,yshift=6] {Slice $n+1$} (-6.5,4);
\draw [decoration={brace,amplitude=7},decorate] (0,4)     -- node [above,yshift=6] {Slice $n$} (9,4);
\begin{scope}[shift={(-5,0)}]
  \filldraw[draw=black,fill=red!10!white,rounded corners=10] (-.5-3.5,-.5) rectangle (4+5,3.5);
  \node at (3.5/2,3) {Compute};
  \filldraw[draw=black,fill=green!50!white] (0,1.5) rectangle (3.5,1+1.5);
  \filldraw[draw=black,fill=blue!30!white] (.5,0) rectangle (2.5+.5,1);
  \pgfmathsetseed{1138};
  \foreach \y in {1/3,2/3}
    \foreach \i in {0,...,3}
      \foreach \j in {0,1}{
        \pgfmathrnd \pgfmathsetmacro{\v}{\pgfmathresult};
        \draw[->] (2.5*\v+.5,\y) -- (3.5*\i/4+3.5*.5/4,\y+1.5);
      }
\end{scope}
\draw[->,very thick] (3.5-5,1.5+.5) -- (0,1.5);
\fill[white] (0,0) rectangle (3.5,2.5);
\fill[green!50!white] (0,1) rectangle (3.5,2);
\node at (3.5/2,3) {Scatter};
\draw[step=1] (0,0) grid (3.5,2.5);
\draw (0,2.5) -- (3.5,2.5) -- (3.5,0);
\foreach \x in {.5,2.5,3.25}{
  \draw[->,very thick] (\x,1) -- (\x,1-1.75);
  \node at (\x,1-1.75-.125) {$\vdots$};
}
\begin{scope}[shift={(-8.5,0)}]
  \fill[white] (0,0) rectangle (1,2.5);
  \fill[blue!30!white] (1,0) rectangle (2,2.5);
  \draw[->,very thick] (2,1.25) -- node [above] {$\curvearrowright$} node [below,sloped] {rotate} (3.5+.5,.5);
  \node at (2/2,3) {Gather};
  \draw[step=1] (0,0) grid (2,2.5);
  \draw (0,2.5) -- (2,2.5);
\end{scope}
\pgfmathsetmacro{\sx}{3.75}
\pgfmathsetmacro{\sy}{.4}
\begin{scope}[shift={(-\sx-8.5,-\sy)}]
  \begin{scope}[shift={(0,2.5)}]
    \filldraw[draw=black,fill=red!10!white,rounded corners=5] (-.25,-.25) rectangle (1.25,1.5) ;
    \node at (.5,1) {Input};
    \filldraw[draw=black,fill=blue!30!white] (0,0) rectangle +(1,.5);
  \end{scope}
  \begin{scope}[shift={(1,0)}]
    \filldraw[draw=black,fill=red!10!white,rounded corners=5] (-.25,-.25) rectangle (1.25,2);
    \node at (.5,1.5) {Input};
    \filldraw[draw=black,fill=blue!30!white] (0,0) rectangle +(1,1);
  \end{scope}
  \begin{scope}[shift={(-1,0)}]
    \filldraw[draw=black,fill=red!10!white,rounded corners=5] (-.25,-.25) rectangle (1.25,2);
    \node at (.5,1.5) {Input};
    \filldraw[draw=black,fill=blue!30!white] (0,0) rectangle +(1,1);
  \end{scope}
  \draw[->,very thick] (1-1,.5) -- (1-1+\sx+1,1+.5+\sy);
  \draw[->,very thick] (1+1,.5) -- (1-1+\sx+1,.5+\sy);
  \draw[->,very thick] (1,.5+2.25) -- (1-1+\sx+1,2+.25+\sy);
\end{scope}
\pgfmathsetmacro{\sx}{4.5}
\pgfmathsetmacro{\sy}{0}
\begin{scope}[shift={(\sx,\sy)}]
  \begin{scope}[shift={(1,0)}]
    \filldraw[draw=black,fill=red!10!white,rounded corners=10] (-.5,-.5) rectangle (4,3.5);
    \node at (3.5/2,3) {Combine};
    \fill[white] (0,0) rectangle (3.5,2.5);
    \fill[green!50!black] (1,1) rectangle (2,2);
    \draw[step=1] (0,0) grid (3.5,2.5);
    \draw (0,2.5) -- (3.5,2.5) -- (3.5,0);
  \end{scope}
  \draw[->,very thick] (2-\sx,1.5) -- (2,1.5);
  \draw[->,very thick] (2.5,1-1.75) -- (2.5,1);
  \node at (2.5,1-1.75-.175) {$\vdots$};
\end{scope}
\end{tikzpicture}
\end{center}
\vspace{-.2in}
\caption{We decompose the set of pentago position into sections, each a 4D array of blocks (shown here as
2D).  The results for a given block are the combination of results from each block line that contains it,
with each such block line depending on exactly one block line from a different section.  Each computation
from input line to output line can be performed on a different processor, first gathering the input blocks
together into a complete line, and finally scattering the output blocks to their owners.  Each pink rounded
rectangle lies in a possibly different process.  Since we compute only those sections which
are unique with symmetries removed, some input lines must be rotated before computation.}
\label{lines}
\vspace{-.1in}
\end{figure*}
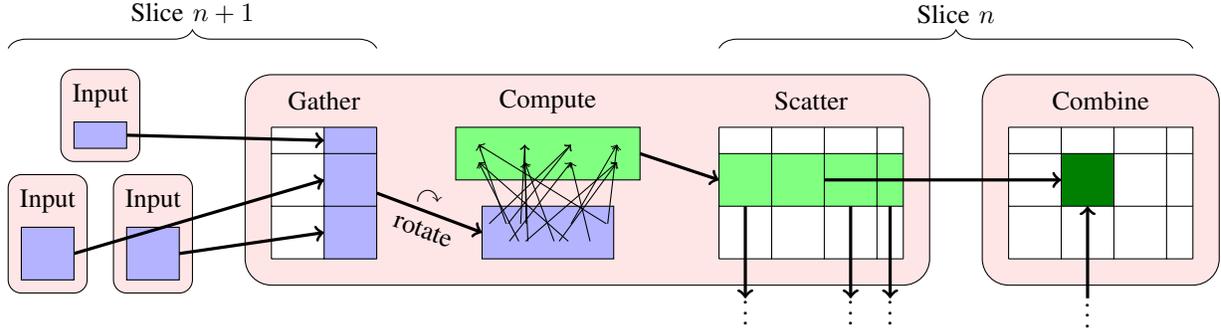

\section{Abstracting over rotations}

The exact symmetry group of pentago is the 8 element dihedral group $D_4$ with 4 global reflections and
4 global rotations.  Computing only one element from each $D_4$ equivalence class saves a factor of $8$,
but does nothing for the large branching factor of the game.  Thus, we consider the \emph{local} group of
all 256 ways to rotate the four quadrants, which has the abelian group structure
$L = \Z_4 \times \Z_4 \times \Z_4 \times \Z_4$.
Combined with the group of global symmetries, the full group of \emph{approximate symmetries} is a semidirect product
$G = \Z_4^4 \rtimes D_4$ with $2048$ elements.

Computing one board $b$ from each equivalence class \wrt $G$ is not enough; we must compute a function
$$f_b : L \to \{-1,0,1\}$$
mapping $g \in L$ to the result of the quadrant rotated board $gb$.  Each board is a win,
loss, or tie, so there are $3^{256}$ such functions.  To avoid ternary arithmetic we use 2 bits per value for
uncompressed data: one bit for win vs. loss/tie and one for win/tie vs. loss.  Thus, for each board we have two
functions $\Z_4^4 \to \{0,1\}$, each a $4 \times 4 \times 4 \times 4$ array of bits.  Each such function
$L \to \{0,1\}$ is packed into a 256 bit table.

Since quadrant rotations do not change the equivalence class \wrt $G$, operating on these functions $f_b$ removes
the branching factor due to rotations.  In its place, we have the mixing operation
\begin{align*}
\rmax &: \left(L \to \{0,1\}\right) \to \left(L \to \{0,1\}\right) \\
\rmax&(f)(g) = \max_{r \in R} f(g+r)
\end{align*}
where $R \subset L$ is the set of $90^\circ$ degree rotations left or right, and we use $+$ because the
group $L$ is abelian.  In addition to $\rmax$, two other rotation abstracted routines are needed.
First, given the position of stones of one color, we must be able to compute the set of rotations
$g \in L$ which produce five in a row.  Second, our equivalence class representative \wrt $G$ can change
when we add a stone, so we must be able to transform $f_b$ into $f_{gb}$ for any $g \in G$; this
involves cyclic shifts, dimension transpositions, and reflections of $4 \times 4 \times 4 \times 4$ bit tables.

Although the code required for these operations is complex, verifying their correctness was a
straightforward process of checking group theoretic definitions against the much simpler routines
operating on one board at a time.  The ease of verification frees us to make the routines as complicated as required
for speed without reducing confidence in the code.

\section{Data layout and distribution}

Given a board $b$, we must choose a unique representative out of the equivalence class $Gb$.  This choice
should be made such that adding a stone changes the representative choice in as few ways as possible, so
that the effective branching factor will be smaller once we take data layout into account.
Concretely, since we have eliminated
branching factor due to rotation, an average board has $12.2$ child boards which are needed as input; if
representatives were chosen arbitrarily, the representatives of the child equivalent classes \wrt $G$
might be located in up to $12.2$ processes depending on how data is distributed.

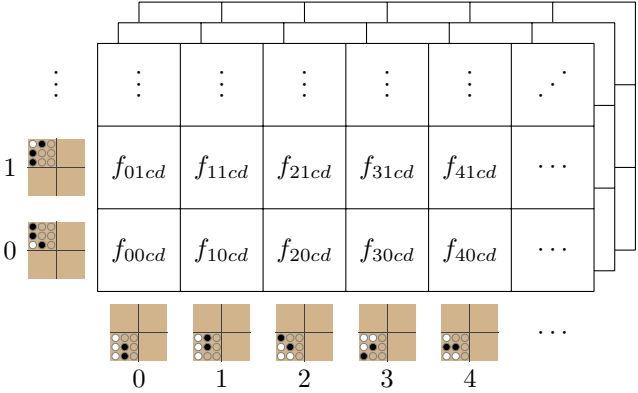
\begin{figure}
\begin{center}
\begin{tikzpicture}[scale=1.1]
  \definecolor{tan}{rgb}{.8235,.7059,.549}
  \begin{scope}[shift={(0.5,0.5)}]
    \fill[white] (0,0) rectangle (6,3);
    \draw[step=1] (0,0) grid (6,3);
  \end{scope}
  \begin{scope}[shift={(0.25,0.25)}]
    \fill[white] (0,0) rectangle (6,3);
    \draw[step=1] (0,0) grid (6,3);
  \end{scope}
  \begin{scope}[shift={(0,0)}]
    \fill[white] (0,0) rectangle (6,3);
    \draw[step=1] (0,0) grid (6,3);
  \end{scope}
  \node at (0.5,-1.05555555556) {$0$};
  \begin{scope}[shift={(0.5,-0.5)},scale=1/9]
    \fill[tan] (-3.05,-3.05) rectangle (3.05,3.05);
    \fill[darkgray] (-0.05,-3.15) rectangle (0.05,3.15);
    \fill[darkgray] (-3.15,-0.05) rectangle (3.15,0.05);
    \filldraw[draw=gray,fill=white,very thin] (-2.55,-2.55) circle (0.4);
    \filldraw[draw=gray,fill=white,very thin] (-2.55,-1.55) circle (0.4);
    \filldraw[draw=gray,fill=white,very thin] (-2.55,-0.55) circle (0.4);
    \filldraw[draw=gray,fill=black,very thin] (-1.55,-2.55) circle (0.4);
    \filldraw[draw=gray,fill=black,very thin] (-1.55,-1.55) circle (0.4);
    \filldraw[draw=gray,fill=tan,very thin] (-1.55,-0.55) circle (0.4);
    \filldraw[draw=gray,fill=tan,very thin] (-0.55,-2.55) circle (0.4);
    \filldraw[draw=gray,fill=tan,very thin] (-0.55,-1.55) circle (0.4);
    \filldraw[draw=gray,fill=tan,very thin] (-0.55,-0.55) circle (0.4);
  \end{scope}
  \node at (1.5,-1.05555555556) {$1$};
  \begin{scope}[shift={(1.5,-0.5)},scale=1/9]
    \fill[tan] (-3.05,-3.05) rectangle (3.05,3.05);
    \fill[darkgray] (-0.05,-3.15) rectangle (0.05,3.15);
    \fill[darkgray] (-3.15,-0.05) rectangle (3.15,0.05);
    \filldraw[draw=gray,fill=white,very thin] (-2.55,-2.55) circle (0.4);
    \filldraw[draw=gray,fill=white,very thin] (-2.55,-1.55) circle (0.4);
    \filldraw[draw=gray,fill=white,very thin] (-2.55,-0.55) circle (0.4);
    \filldraw[draw=gray,fill=tan,very thin] (-1.55,-2.55) circle (0.4);
    \filldraw[draw=gray,fill=black,very thin] (-1.55,-1.55) circle (0.4);
    \filldraw[draw=gray,fill=black,very thin] (-1.55,-0.55) circle (0.4);
    \filldraw[draw=gray,fill=tan,very thin] (-0.55,-2.55) circle (0.4);
    \filldraw[draw=gray,fill=tan,very thin] (-0.55,-1.55) circle (0.4);
    \filldraw[draw=gray,fill=tan,very thin] (-0.55,-0.55) circle (0.4);
  \end{scope}
  \node at (2.5,-1.05555555556) {$2$};
  \begin{scope}[shift={(2.5,-0.5)},scale=1/9]
    \fill[tan] (-3.05,-3.05) rectangle (3.05,3.05);
    \fill[darkgray] (-0.05,-3.15) rectangle (0.05,3.15);
    \fill[darkgray] (-3.15,-0.05) rectangle (3.15,0.05);
    \filldraw[draw=gray,fill=white,very thin] (-2.55,-2.55) circle (0.4);
    \filldraw[draw=gray,fill=white,very thin] (-2.55,-1.55) circle (0.4);
    \filldraw[draw=gray,fill=black,very thin] (-2.55,-0.55) circle (0.4);
    \filldraw[draw=gray,fill=white,very thin] (-1.55,-2.55) circle (0.4);
    \filldraw[draw=gray,fill=black,very thin] (-1.55,-1.55) circle (0.4);
    \filldraw[draw=gray,fill=tan,very thin] (-1.55,-0.55) circle (0.4);
    \filldraw[draw=gray,fill=tan,very thin] (-0.55,-2.55) circle (0.4);
    \filldraw[draw=gray,fill=tan,very thin] (-0.55,-1.55) circle (0.4);
    \filldraw[draw=gray,fill=tan,very thin] (-0.55,-0.55) circle (0.4);
  \end{scope}
  \node at (3.5,-1.05555555556) {$3$};
  \begin{scope}[shift={(3.5,-0.5)},scale=1/9]
    \fill[tan] (-3.05,-3.05) rectangle (3.05,3.05);
    \fill[darkgray] (-0.05,-3.15) rectangle (0.05,3.15);
    \fill[darkgray] (-3.15,-0.05) rectangle (3.15,0.05);
    \filldraw[draw=gray,fill=black,very thin] (-2.55,-2.55) circle (0.4);
    \filldraw[draw=gray,fill=white,very thin] (-2.55,-1.55) circle (0.4);
    \filldraw[draw=gray,fill=white,very thin] (-2.55,-0.55) circle (0.4);
    \filldraw[draw=gray,fill=tan,very thin] (-1.55,-2.55) circle (0.4);
    \filldraw[draw=gray,fill=black,very thin] (-1.55,-1.55) circle (0.4);
    \filldraw[draw=gray,fill=white,very thin] (-1.55,-0.55) circle (0.4);
    \filldraw[draw=gray,fill=tan,very thin] (-0.55,-2.55) circle (0.4);
    \filldraw[draw=gray,fill=tan,very thin] (-0.55,-1.55) circle (0.4);
    \filldraw[draw=gray,fill=tan,very thin] (-0.55,-0.55) circle (0.4);
  \end{scope}
  \node at (4.5,-1.05555555556) {$4$};
  \begin{scope}[shift={(4.5,-0.5)},scale=1/9]
    \fill[tan] (-3.05,-3.05) rectangle (3.05,3.05);
    \fill[darkgray] (-0.05,-3.15) rectangle (0.05,3.15);
    \fill[darkgray] (-3.15,-0.05) rectangle (3.15,0.05);
    \filldraw[draw=gray,fill=white,very thin] (-2.55,-2.55) circle (0.4);
    \filldraw[draw=gray,fill=black,very thin] (-2.55,-1.55) circle (0.4);
    \filldraw[draw=gray,fill=white,very thin] (-2.55,-0.55) circle (0.4);
    \filldraw[draw=gray,fill=white,very thin] (-1.55,-2.55) circle (0.4);
    \filldraw[draw=gray,fill=black,very thin] (-1.55,-1.55) circle (0.4);
    \filldraw[draw=gray,fill=tan,very thin] (-1.55,-0.55) circle (0.4);
    \filldraw[draw=gray,fill=tan,very thin] (-0.55,-2.55) circle (0.4);
    \filldraw[draw=gray,fill=tan,very thin] (-0.55,-1.55) circle (0.4);
    \filldraw[draw=gray,fill=tan,very thin] (-0.55,-0.55) circle (0.4);
  \end{scope}
  \node at (5.35714285714,-0.5) {$\cdot$};
  \node at (5.5,-0.5) {$\cdot$};
  \node at (5.64285714286,-0.5) {$\cdot$};
  \node at (-1.05555555556,0.5) {$0$};
  \begin{scope}[shift={(-0.5,0.5)},scale=1/9]
    \fill[tan] (-3.05,-3.05) rectangle (3.05,3.05);
    \fill[darkgray] (-0.05,-3.15) rectangle (0.05,3.15);
    \fill[darkgray] (-3.15,-0.05) rectangle (3.15,0.05);
    \filldraw[draw=gray,fill=white,very thin] (-2.55,0.55) circle (0.4);
    \filldraw[draw=gray,fill=black,very thin] (-2.55,1.55) circle (0.4);
    \filldraw[draw=gray,fill=black,very thin] (-2.55,2.55) circle (0.4);
    \filldraw[draw=gray,fill=black,very thin] (-1.55,0.55) circle (0.4);
    \filldraw[draw=gray,fill=tan,very thin] (-1.55,1.55) circle (0.4);
    \filldraw[draw=gray,fill=tan,very thin] (-1.55,2.55) circle (0.4);
    \filldraw[draw=gray,fill=tan,very thin] (-0.55,0.55) circle (0.4);
    \filldraw[draw=gray,fill=tan,very thin] (-0.55,1.55) circle (0.4);
    \filldraw[draw=gray,fill=tan,very thin] (-0.55,2.55) circle (0.4);
  \end{scope}
  \node at (-1.05555555556,1.5) {$1$};
  \begin{scope}[shift={(-0.5,1.5)},scale=1/9]
    \fill[tan] (-3.05,-3.05) rectangle (3.05,3.05);
    \fill[darkgray] (-0.05,-3.15) rectangle (0.05,3.15);
    \fill[darkgray] (-3.15,-0.05) rectangle (3.15,0.05);
    \filldraw[draw=gray,fill=black,very thin] (-2.55,0.55) circle (0.4);
    \filldraw[draw=gray,fill=black,very thin] (-2.55,1.55) circle (0.4);
    \filldraw[draw=gray,fill=white,very thin] (-2.55,2.55) circle (0.4);
    \filldraw[draw=gray,fill=tan,very thin] (-1.55,0.55) circle (0.4);
    \filldraw[draw=gray,fill=tan,very thin] (-1.55,1.55) circle (0.4);
    \filldraw[draw=gray,fill=black,very thin] (-1.55,2.55) circle (0.4);
    \filldraw[draw=gray,fill=tan,very thin] (-0.55,0.55) circle (0.4);
    \filldraw[draw=gray,fill=tan,very thin] (-0.55,1.55) circle (0.4);
    \filldraw[draw=gray,fill=tan,very thin] (-0.55,2.55) circle (0.4);
  \end{scope}
  \node at (-0.5,2.35714285714) {$\cdot$};
  \node at (-0.5,2.5) {$\cdot$};
  \node at (-0.5,2.64285714286) {$\cdot$};
  \node at (.5+0,.5+0) {$f_{00cd}$};
  \node at (.5+0,.5+1) {$f_{01cd}$};
  \node at (0.5,2.35714285714) {$\cdot$};
  \node at (0.5,2.5) {$\cdot$};
  \node at (0.5,2.64285714286) {$\cdot$};
  \node at (.5+1,.5+0) {$f_{10cd}$};
  \node at (.5+1,.5+1) {$f_{11cd}$};
  \node at (1.5,2.35714285714) {$\cdot$};
  \node at (1.5,2.5) {$\cdot$};
  \node at (1.5,2.64285714286) {$\cdot$};
  \node at (.5+2,.5+0) {$f_{20cd}$};
  \node at (.5+2,.5+1) {$f_{21cd}$};
  \node at (2.5,2.35714285714) {$\cdot$};
  \node at (2.5,2.5) {$\cdot$};
  \node at (2.5,2.64285714286) {$\cdot$};
  \node at (.5+3,.5+0) {$f_{30cd}$};
  \node at (.5+3,.5+1) {$f_{31cd}$};
  \node at (3.5,2.35714285714) {$\cdot$};
  \node at (3.5,2.5) {$\cdot$};
  \node at (3.5,2.64285714286) {$\cdot$};
  \node at (.5+4,.5+0) {$f_{40cd}$};
  \node at (.5+4,.5+1) {$f_{41cd}$};
  \node at (4.5,2.35714285714) {$\cdot$};
  \node at (4.5,2.5) {$\cdot$};
  \node at (4.5,2.64285714286) {$\cdot$};
  \node at (5.35714285714,0.5) {$\cdot$};
  \node at (5.5,0.5) {$\cdot$};
  \node at (5.64285714286,0.5) {$\cdot$};
  \node at (5.35714285714,1.5) {$\cdot$};
  \node at (5.5,1.5) {$\cdot$};
  \node at (5.64285714286,1.5) {$\cdot$};
  \node at (5.35714285714,2.35714285714) {$\cdot$};
  \node at (5.5,2.5) {$\cdot$};
  \node at (5.64285714286,2.64285714286) {$\cdot$};
\end{tikzpicture}
\end{center}
\vspace{-.15in}
\caption{Each section is a 4D array of functions $f_{abcd} : L \to \{-1,0,1\}$.  Each dimension
corresponds to all patterns of stones in one of the four quadrants with fixed counts of black and white
stones, including only patterns lexicographically minimal under rotation.  For each $(a,b,c,d)$
describing the four quadrants of a board, $f_{abcd}$ gives the loss/tie/win values for all 256 ways to rotate the
four quadrants.  The order is chosen so that reflected pairs are adjacent so that
reflection preserves the block structure.  The figure shows 2D slices of the full 4D array.}
\label{section}
\vspace{-.1in}
\end{figure}

Therefore, we partition all boards in a given slice (fixed number of stones) into \emph{sections} defined by the
numbers of stones of each color in the four quadrants, computing only sections whose counts are lexicographically
minimal under $D_4$ symmetry.  Within a given section, we consider only boards whose quadrants are lexicographically
minimal under per-quadrant rotations; each quadrant is independent under this requirement, so the section becomes
a four dimensional rectangular array where each dimension defines the stones in one quadrant.  We precompute an
ordering of these rotation minimal quadrant states so that we can convert from a position in the four dimensional
section array to a board state using table lookup.  The structure of one such section is shown in \autoref{section}.
There are at most four quadrants to chose from when placing a stone (some may be full near the end of the game),
so at most four child sections contribute to the results for a given parent.  In other words, we have reduced the
effective branching factor from $12.2$ to $4$.

Sections alone provide insufficiently fine parallelism (the largest is $1.8$ TB uncompressed), so we
divide each 4D section array into $8 \times 8 \times 8 \times 8$ blocks and partition the blocks for all
sections among the different processes.  When a stone is added in a quadrant, we move to a child section with
index layout different from the parent only for the quadrant where the stone was added
since the other quadrants have the same pattern of stones.  Therefore, a single block in a parent section depends
on inputs from one \emph{line} of blocks in up to four child sections.  Since the different input lines for a block
correspond to moves in different quadrants, we can compute each line contribution separately on different processes
and combine them with $\max$ on whichever process owns the output block.

The structure of the computation is illustrated in \autoref{lines}.  Say we want to compute an output block line
in section $n$, which is a $k \times 1 \times 1 \times 1$ grid of blocks (possibly transposed) corresponding to
an $8k \times 8 \times 8 \times 5$ grid of boards (block sizes may differ from $8$ at section boundaries).
Our output block line depends on a single input block line in section $n+1$, which (possibly after rotation) is an
$8k' \times 8 \times 8 \times 5$ grid of nodes.  The input and output block lines differ in size only along the
long dimension, since the long dimension corresponds to the quadrant where we will place a stone.  When we
compute index $(a,b,c,d)$ of the output line, we mix together several indices $(a',b,c,d)$ with different $a'$
corresponding to the different places to put a stone in the lines' quadrant.  Since the map from
$a$ to $a'$ is many to many, computing the entire block line on a single processor gives an effective
branching factor of $4$ for communication cost even though the underlying branching factor is $12.2$.
Once the block line is computed, its component blocks are scattered to their owners to be merged together
via $\max$ with other contributions (each block needs up to four such block line contributions).

Since our block structure is symmetric \wrt dimension and our quadrant positions are always minimal with respect
to rotation, the block structure of sections is preserved when the board is rotated.  However, the ordering of
quadrant states does change when a quadrant is reflected, since a lexicographically least quadrant may no longer
be lexicographically least after reflection.  To maintain the block structure, we require even sized blocks
(8 in our case) and adjust our precomputed quadrant state ordering so that reflected pairs occur next to each
other in the same block.  With this trick, the block structure is invariant to all symmetries.

The relatively simple structure of sections, blocks, and rotation-abstracted values within blocks does have
a cost: if a position or section is preserved by a symmetry it will be double counted in the data layout.
Abstracting over rotations increases the number of effective positions by 5.4\% and removing symmetries only
at the section level costs an additional 9.3\%, for a total overcounting of 15.2\% relative to storing each
symmetry-unique position once.\cprotect\footnote{Run \verb+analyze ratio+.}

\subsection{Deterministic pseudorandom partitioning}

In parallel, we must partition the set of blocks across processes to balance memory usage and the
set of block lines across processes to balance compute.  Ideally, the process computing a given block line
would also own many of the input and output blocks in order to minimize communication.  Unfortunately, these
desires couple together the partition for all slices.  Over all slices, there are 3,654,002,393 blocks and
996,084,744 block lines.\cprotect\footnote{Run \verb+analyze approx+.}  Thus, we have a graph partitioning
problem with 4,650,087,137 nodes divided into 72 clusters, each cluster defining a load balancing constraint.
Although existing graph partitioning codes such as ParMETIS\cite{schloegel2002parmetis} might be sufficient
for our problem, we have sufficient computation to hide communication latency and opt for a
simple randomized partitioning scheme instead.

We partition each slice independently.  Since there are at most 8239 sections to a slice, and
each section is a regular 4D grid of blocks, we can define an ordering of all block lines
by arranging the sections back to back.  We choose a pseudorandom permutation of the ordered block
lines and give each process a contiguous chunk of the scrambled ordering.  Each block is then randomly
assigned to one of its four lines, and given to the process which owns that line.  In both cases, these choices
can be made consistently with only an $O(1)$ size random seed shared between processes: we use the arbitrary size
cipher technique of \cite{black2002ciphers} for random access random permutations and the Threefry generator
of \cite{salmon2011parallel} for conventional random numbers.  Since the cipher permutations are invertible,
we can find the process owning a given block or block line in $O(1)$ time.

\begin{figure}
\begin{center}
\includegraphics[width=\columnwidth]{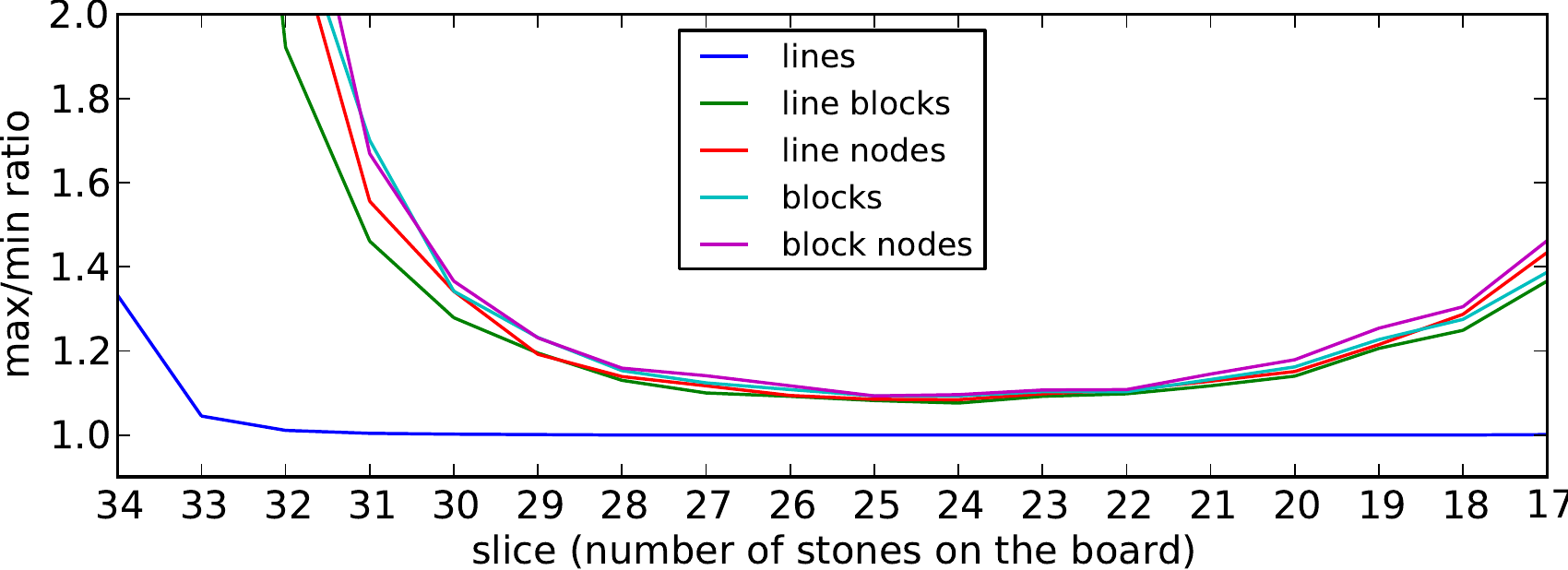}
\end{center}
\vspace{-.1in}
\cprotect\caption{Load balance ratio ($\max/\min$) for various quantities as a function of stones on the board using
deterministic pseudorandom partitioning.  89\% of the computation occurs from slice $20$ to $28$, where
all quantities balance to within 20\%.  Run \verb+paper/numbers load+ to reproduce.}
\label{balance}
\vspace{-.1in}
\end{figure}

At scale, a pseudorandom partitioning scheme automatically balances any quantity where the central limit
theorem applies.  In particular, though our scheme does not explicitly account for the different amounts of
work required to compute different block lines, or the different sizes of blocks at the boundary of sections,
there are enough blocks and block lines to keep the $\max/\min$ ratio to within 10-20\% for all large slices and
all relevant quantities (\autoref{balance}).

\subsection{Compression}

Since our uncompressed memory usage would be at least $213$ TB, we compress all data in memory until needed
using the fast but weak compression library Snappy \cite{snappy2014}.  Most blocks are $256$ KB
($64 \cdot 8^4$ bytes) uncompressed, large enough to compress each block separately without harming compression ratio.
Despite its speed relative to stronger compression such as ZLIB or LZMA\cite{deutsch1996zlib,xz2014}, Snappy still consumed about 29\% of our
compute time ignoring I/O.\cprotect\footnote{See \verb+snappy fraction+ in \verb+paper/numbers+.}  Stronger
compression is thus out of reach for in memory purposes, although we do use LZMA when writing
out the smaller final data set.

With compression the memory usage varies unpredictably, with two consequences.  First, repeatedly
allocating and deallocating irregular block sizes results in significant fragmentation.  During early testing on
BlueGene, which has no virtual memory system, fragmentation caused the code to run out of memory much earlier
than necessary.  We solved this with a manual compacting garbage collector for bulk data storage, which is
straightforward in our case due to the lack of pointers.  Second, estimates from slices near the end of the game
gave a compression ratio of roughly $1/3$.  Since we were uncertain whether this ratio should grow or shrink at
the peak of the computation, and wanted a high probability of solving the game in a single run, we used a
conservative estimate of $0.4$ when determining how many nodes to use.  However, the actual average compression
ratio was $0.26$.\cprotect\footnote{See \verb+total data+ in \verb+paper/numbers+.}  Taking advantage
of the unexpectedly good compression would have required dynamic partitioning, or even (ideally) a dynamic
number of MPI nodes.

\section{Asynchronous control flow}

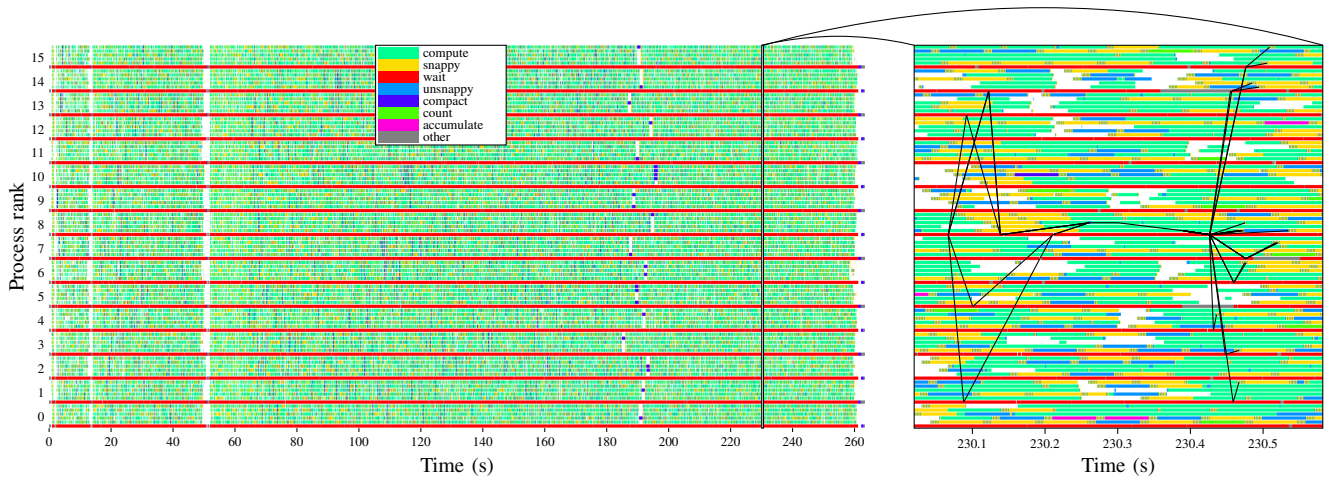
\begin{figure*}
\begin{center}
\begin{tikzpicture}
  \node[anchor=south west,inner sep=0] (full) at (0,0) {\includegraphics[height=2.0in]{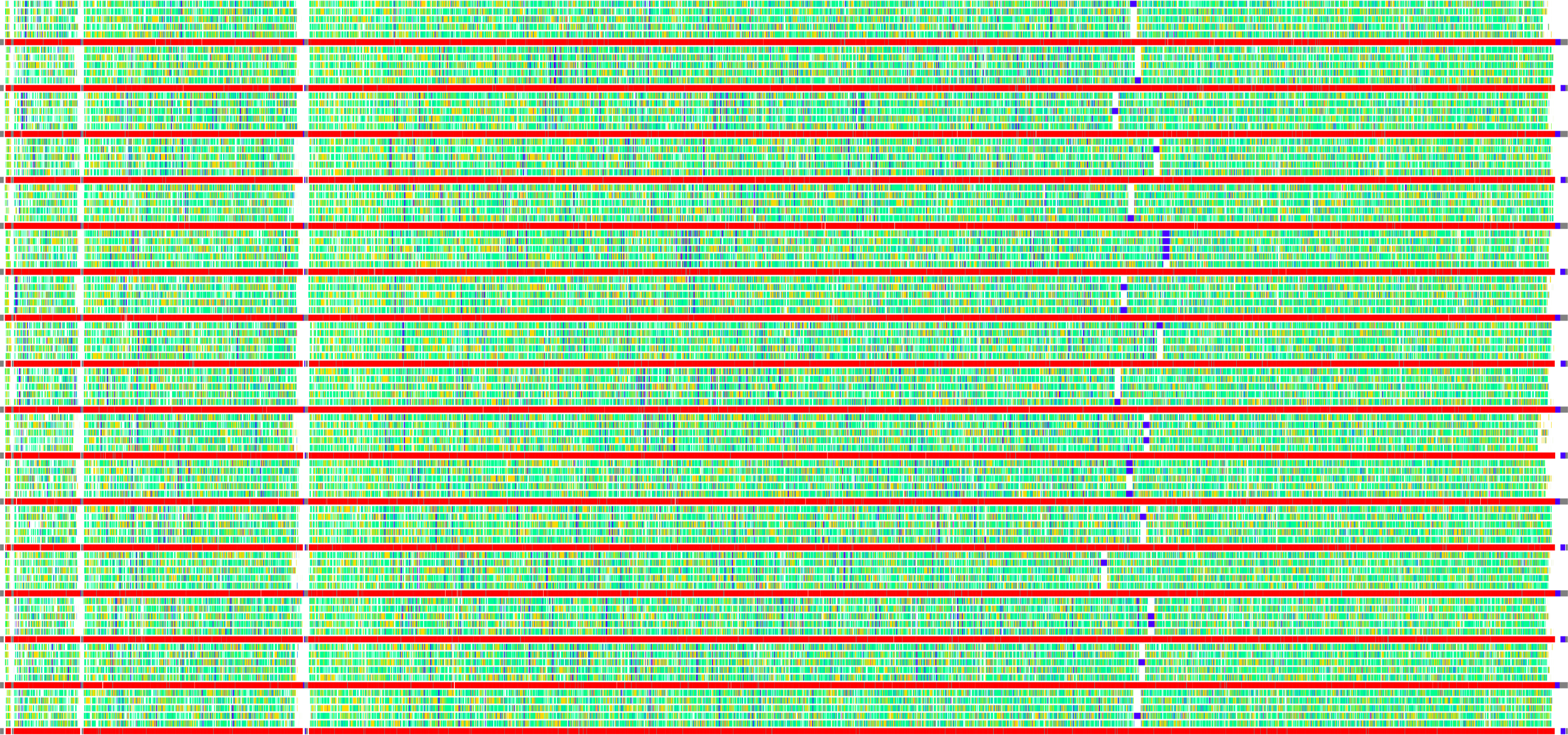}};
  \begin{scope}[x={(full.south east)},y={(full.north west)}]
    \foreach \i in {0,...,15}{
      \node at (0,\i/16+1/32) [left,scale=.5] {\i};
    }
    \node at (-.04,.5) [rotate=90,scale=.8] {Process rank}; 
    \foreach \j in {0,20,...,260}{
      \draw (\j/263.473,0) -- +(0,-.01);
      \node at (\j/263.473,-.01) [below,scale=.5] {\j};
    }
    \node at (.5,-.1) [scale=.8] {Time (s)}; 
    \begin{scope}[shift={(.4,0)}]
      \filldraw[draw=black,fill=white] (0,1) rectangle +(0.16,-0.260417);
      \definecolor{compute}{rgb}{0,1,0.571429};
      \fill[compute] (0.0025,1-0.00520833) rectangle +(0.05,-0.03125);
      \node [right,scale=.5,anchor=base west] at (0.0525,1-0.0302083) {compute};
      \definecolor{snappy}{rgb}{1,0.857143,0};
      \fill[snappy] (0.0025,1-0.0364583) rectangle +(0.05,-0.03125);
      \node [right,scale=.5,anchor=base west] at (0.0525,1-0.0614583) {snappy};
      \definecolor{wait}{rgb}{1,0,0};
      \fill[wait] (0.0025,1-0.0677083) rectangle +(0.05,-0.03125);
      \node [right,scale=.5,anchor=base west] at (0.0525,1-0.0927083) {wait};
      \definecolor{unsnappy}{rgb}{0,0.571429,1};
      \fill[unsnappy] (0.0025,1-0.0989583) rectangle +(0.05,-0.03125);
      \node [right,scale=.5,anchor=base west] at (0.0525,1-0.123958) {unsnappy};
      \definecolor{compact}{rgb}{0.285714,0,1};
      \fill[compact] (0.0025,1-0.130208) rectangle +(0.05,-0.03125);
      \node [right,scale=.5,anchor=base west] at (0.0525,1-0.155208) {compact};
      \definecolor{count}{rgb}{0.285714,1,0};
      \fill[count] (0.0025,1-0.161458) rectangle +(0.05,-0.03125);
      \node [right,scale=.5,anchor=base west] at (0.0525,1-0.186458) {count};
      \definecolor{accumulate}{rgb}{1,0,0.857143};
      \fill[accumulate] (0.0025,1-0.192708) rectangle +(0.05,-0.03125);
      \node [right,scale=.5,anchor=base west] at (0.0525,1-0.217708) {accumulate};
      \definecolor{other}{rgb}{0.5,0.5,0.5};
      \fill[other] (0.0025,1-0.223958) rectangle +(0.05,-0.03125);
      \node [right,scale=.5,anchor=base west] at (0.0525,1-0.248958) {other};
    \end{scope}
    \filldraw[draw=black,fill=white] (0.8730960728423784,0) rectangle (0.8752302148607257,1);
    \coordinate (fromleft) at (0.8730960728423784,1);
    \coordinate (fromright) at (0.8752302148607257,1);
  \end{scope}
  \begin{scope}[shift={(11.5,0)}]
    \node[anchor=south west,inner sep=0] (image) at (0,0) {\includegraphics[height=2.0in]{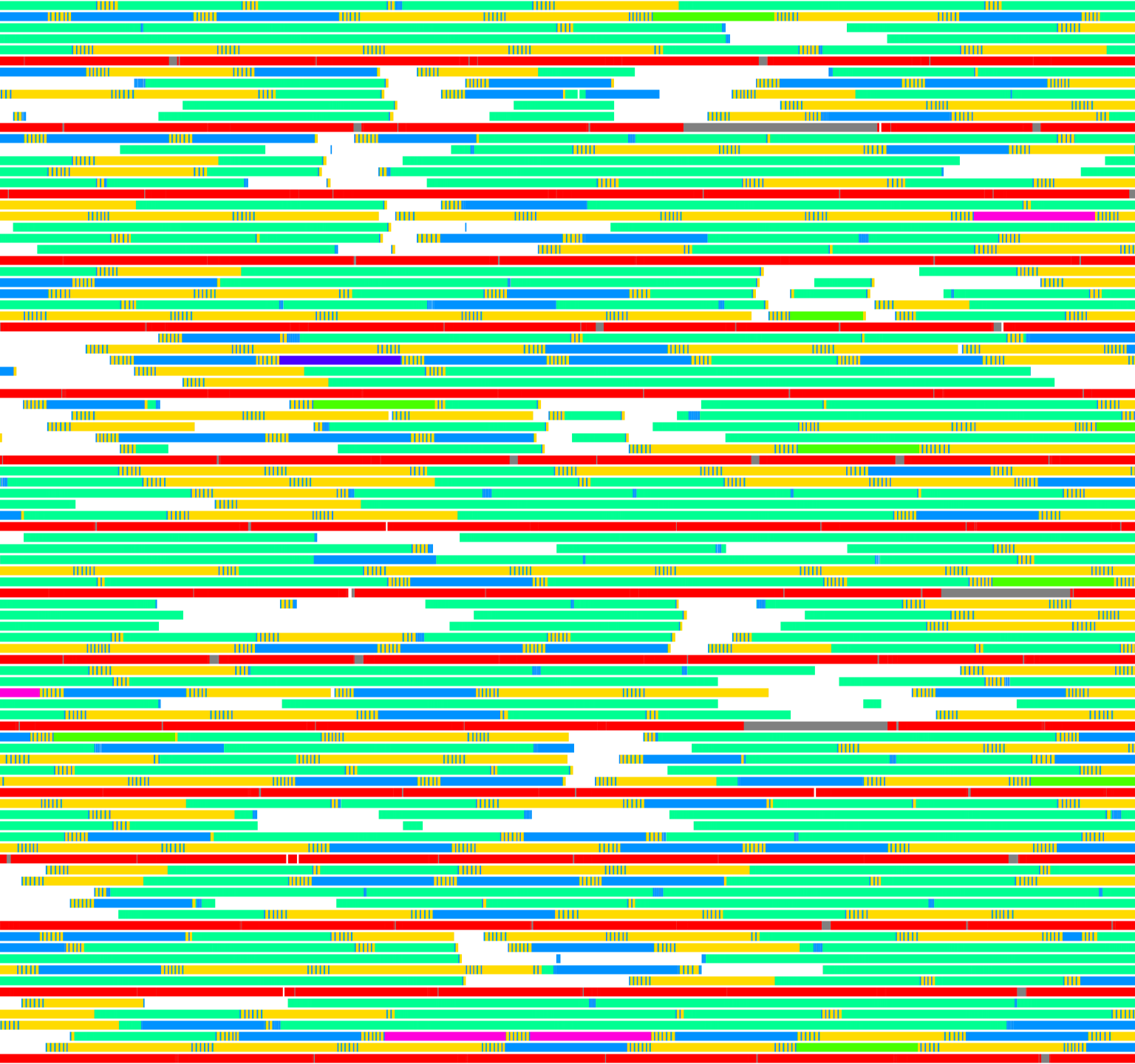}};
    \begin{scope}[x={(image.south east)},y={(image.north west)}]
      \draw (0,0) rectangle (1,1);
      \draw (0.498646,0.536458) -- (0.428306,0.536458); \draw (0.428306,0.536458) -- (0.211131,0.505208); \draw (0.428306,0.536458) -- (0.339567,0.505208); \draw (0.428306,0.536458) -- (0.338751,0.505208); \draw (0.428306,0.536458) -- (0.210792,0.505208); \draw (0.428306,0.536458) -- (0.211169,0.505208); \draw (0.211169,0.505208) -- (0.18248,0.880208); \draw (0.18248,0.880208) -- (0.0833778,0.505208); \draw (0.0833778,0.505208) -- (0.0833333,0.505208); \draw (0.210792,0.505208) -- (0.128093,0.817708); \draw (0.128093,0.817708) -- (0.0833716,0.505208); \draw (0.0833716,0.505208) -- (0.0833333,0.505208); \draw (0.338751,0.505208) -- (0.121249,0.0677083); \draw (0.121249,0.0677083) -- (0.0833653,0.505208); \draw (0.0833653,0.505208) -- (0.0833333,0.505208); \draw (0.339567,0.505208) -- (0.143496,0.317708); \draw (0.143496,0.317708) -- (0.0833556,0.505208); \draw (0.0833556,0.505208) -- (0.0833333,0.505208); \draw (0.211131,0.505208) -- (0.182467,0.880208); \draw (0.182467,0.880208) -- (0.0833449,0.505208); \draw (0.0833449,0.505208) -- (0.0833333,0.505208); \draw (0.498646,0.536458) -- (0.658716,0.515625); \draw (0.658716,0.515625) -- (0.722719,0.505208); \draw (0.658716,0.515625) -- (0.722741,0.505208); \draw (0.658716,0.515625) -- (0.722758,0.505208); \draw (0.658716,0.515625) -- (0.722817,0.505208); \draw (0.658716,0.515625) -- (0.722873,0.505208); \draw (0.658716,0.515625) -- (0.722928,0.505208); \draw (0.658716,0.515625) -- (0.723026,0.505208); \draw (0.658716,0.515625) -- (0.723078,0.505208); \draw (0.658716,0.515625) -- (0.723085,0.505208); \draw (0.658716,0.515625) -- (0.723096,0.505208); \draw (0.658716,0.515625) -- (0.723151,0.505208); \draw (0.658716,0.515625) -- (0.723251,0.505208); \draw (0.658716,0.515625) -- (0.723351,0.505208); \draw (0.658716,0.515625) -- (0.723405,0.505208); \draw (0.658716,0.515625) -- (0.723461,0.505208); \draw (0.658716,0.515625) -- (0.723518,0.505208); \draw (0.658716,0.515625) -- (0.723526,0.505208); \draw (0.658716,0.515625) -- (0.723577,0.505208); \draw (0.658716,0.515625) -- (0.723631,0.505208); \draw (0.658716,0.515625) -- (0.723639,0.505208); \draw (0.723639,0.505208) -- (0.777085,0.880208); \draw (0.777085,0.880208) -- (0.826283,0.901042); \draw (0.723631,0.505208) -- (0.782422,0.380208); \draw (0.782422,0.380208) -- (0.8089,0.432292); \draw (0.723577,0.505208) -- (0.72372,0.505208); \draw (0.72372,0.505208) -- (0.789167,0.515625); \draw (0.723526,0.505208) -- (0.811573,0.942708); \draw (0.811573,0.942708) -- (0.863152,0.953125); \draw (0.723518,0.505208) -- (0.811551,0.442708); \draw (0.811551,0.442708) -- (0.884799,0.484375); \draw (0.723461,0.505208) -- (0.723717,0.505208); \draw (0.723717,0.505208) -- (0.792687,0.515625); \draw (0.723405,0.505208) -- (0.811528,0.442708); \draw (0.811528,0.442708) -- (0.888,0.484375); \draw (0.723351,0.505208) -- (0.723712,0.505208); \draw (0.723712,0.505208) -- (0.795999,0.515625); \draw (0.723251,0.505208) -- (0.723674,0.505208); \draw (0.723674,0.505208) -- (0.810391,0.536458); \draw (0.723151,0.505208) -- (0.7237,0.505208); \draw (0.7237,0.505208) -- (0.806281,0.515625); \draw (0.723096,0.505208) -- (0.811523,0.442708); \draw (0.811523,0.442708) -- (0.891369,0.484375); \draw (0.723085,0.505208) -- (0.782361,0.380208); \draw (0.782361,0.380208) -- (0.812741,0.432292); \draw (0.723078,0.505208) -- (0.774176,0.880208); \draw (0.774176,0.880208) -- (0.843946,0.890625); \draw (0.723026,0.505208) -- (0.723705,0.505208); \draw (0.723705,0.505208) -- (0.802709,0.515625); \draw (0.722928,0.505208) -- (0.723708,0.505208); \draw (0.723708,0.505208) -- (0.799383,0.515625); \draw (0.722873,0.505208) -- (0.780705,0.0677083); \draw (0.780705,0.0677083) -- (0.794775,0.119792); \draw (0.722817,0.505208) -- (0.723695,0.505208); \draw (0.723695,0.505208) -- (0.916667,0.515625); \draw (0.722758,0.505208) -- (0.761403,0.192708); \draw (0.761403,0.192708) -- (0.795137,0.203125); \draw (0.722741,0.505208) -- (0.732447,0.255208); \draw (0.732447,0.255208) -- (0.739917,0.296875); \draw (0.722719,0.505208) -- (0.811442,0.942708); \draw (0.811442,0.942708) -- (0.86977,0.994792);
      \coordinate (toleft) at (0,1);
      \coordinate (toright) at (1,1);
      \foreach \i in {230.1,230.2,230.3,230.4,230.5}{
        \draw (\i/.5622888-409.1087026809,0) -- +(0,-.015);
        \node at (\i/.5622888-409.1087026809,-.015) [below,scale=.5] {\i};
      }
      \node at (.5,-.1) [scale=.8] {Time (s)};
    \end{scope}
  \end{scope}
  \draw plot [smooth,tension=1] coordinates { (fromleft) ($ .5*(fromleft) + .5*(toleft) + (0,.12) $) (toleft) };
  \draw plot [smooth,tension=1] coordinates { (fromright) ($ .5*(fromright) + .5*(toright) + (0,.5) $) (toright) };
\end{tikzpicture}
\end{center}
\vspace{-.15in}
\cprotect\caption{Images from a trace visualization tool used to diagnose performance problems in asynchronous
code.  (Left) The history of a 16 process, 96 thread run computing the section with four stones of
each color in each quadrant.  Each process has one communication thread (mostly red for waiting) and five worker
threads performing computation, with colors showing the type of computation performed.  (Right) A zoom showing the
information flow related to part of a block line computation.  At the leftmost point in the graph shown, the process
decides to compute a given block line, and sends out requests for input data to other processes.
Once all responses arrive, the computation begins.  When the computation finishes, the results are scattered to
other processes and compressed for storage.  To reproduce, run \verb+paper/history+.}
\label{history}
\vspace{-.1in}
\end{figure*}

Within slice $n$, each process can compute its allocation of block lines in any order, since all inputs
are from slice $n+1$ which has already been computed.  However, most of these inputs are stored on other
processes, and due to memory limitations only a small fraction of them can be stored locally at any given time.
Moreover, the time to compute a given block line varies with size, ruling out a lockstep
communication/compute cycle.  Instead, we use an asynchronous control flow where each process sends
requests for input data for at most five block lines at time, begins computing as soon as all inputs for a
block line are in place, sends out output data when ready, and listens for incoming output data from other
processes to be merged.

We emphasize that asynchrony is needed only because of the memory constraint: if we had 4 times as much memory
in order to store all inputs locally, we could split the computation into communication / compute epochs and
use an embarrassingly parallel control flow during compute.

We use a hybrid MPI/Pthread model where each 6 thread process has 1 communication thread and 5 worker threads
(with 8 processes per 48-hyperthread Edison node).
A hybrid structure reduces the memory usage by allowing several threads to share the same temporary storage
required when computing a block line.  The communication thread must simultaneously listen for
incoming remote messages and completed tasks from the worker threads; this can be done with self-to-self MPI
messages in environments which support \verb+MPI_THREAD_MULTIPLE+ but requires alternatively polling between
\verb+MPI_Testsome+ and \verb+pthread_spin_trylock+ if only \verb+MPI_THREAD_FUNNELED+ is available.

At the time the code was written, the MPI 3 standard was not yet available on the target machine, and the
one sided communication primitives in MPI 2 were not sufficient for our communication
pattern.\footnote{For more discussion, see \url{http://scicomp.stackexchange.com/questions/2846/simulating-the-mpi-isend-irecv-wait-model-with-one-sided-communication}.}
Specifically, the MPI 2 one sided primitives provide no asynchronous way to know when a request completes; and
our only synchronization points are between entire slices.  MPI 3 solves this problem:
after an initial communication phase exchanging pointers to the required compressed slice $n+1$ blocks, all
input requests during slice $n$ computation could be handled with \verb+MPI_Rget+\cite{mpi-3:2012}.
Unfortunately, MPI 3 does
not solve the reverse problem of output messages: when an output block arrives, any previous data for that
block must be uncompressed, combined with the new data, and recompressed for storage.  \verb+MPI_Accumulate+ has
no support for user defined operations, so output messages would still be limited to two sided communication.
Finally, MPI 3 provides the useful \verb+MPI_Ibarrier+ primitive which is exactly what we need to know when
all processes have finished computing and thus when the previous slice can be deallocated; since we use MPI 2
we must simulate \verb+MPI_Ibarrier+ using a manual tree reduction.

The asynchronous control flow was tricky to write but straightforward to debug,
since most bugs manifested as deadlocks.  Each message and response is labeled with a unique global id,
so deadlocks were easy to eliminate by reading traces of events.  However, the performance
characteristics of the code were harder to understand, since high latency might be a result of
unrelated communication at the same time.  Existing profiling tools such as TAU
\cite{shende2006tau} were insufficient for tracing the dependencies between asynchronous messages
combined with control flow across threads.  Thus, we wrote a custom trace visualizer with
knowledge of the information flow between inputs through compute to output; an example visualization is
shown in \autoref{history}.  On NERSC's Cray XE6 Hopper, this tool confirmed that long idle periods
were due to high latency, but was not sufficient to diagnose the underlying cause of the problem.
Unfortunately, we still do not know the cause of this latency.  Testing on the Argonne's BlueGene/Q Vesta
was inconclusive since the code easily saturated BlueGene's poor integer performance.  On the newer Cray
Edison used for the final production run, the problem went away: worker threads were idle only
16.4\% of the time with I/O excluded.\cprotect\footnote{See \verb+Idle vs. total time+ in
\verb+paper/numbers+\label{idle}.}

\section{Performance}

Our final production job ran on NERSC's Cray XC30 Edison, using 98304 ($3 \times 2^{15}$) threads
including hyperthreading (49152 cores, 2048 nodes).  The bulk of the computation from slice $35$ down
to $19$ took $2.7$ hours.  Starting at slice $18$ we began writing output results to disk, though our
first computation finished writing only slices $17$ and $18$ before hitting an unfortunately chosen
wall clock limit of $4$ hours.  Two smaller jobs on 192 and 128 nodes were used to finish the computation
down to slice $0$, the start of the game.

\begin{figure}
\begin{center}
\includegraphics[width=\columnwidth]{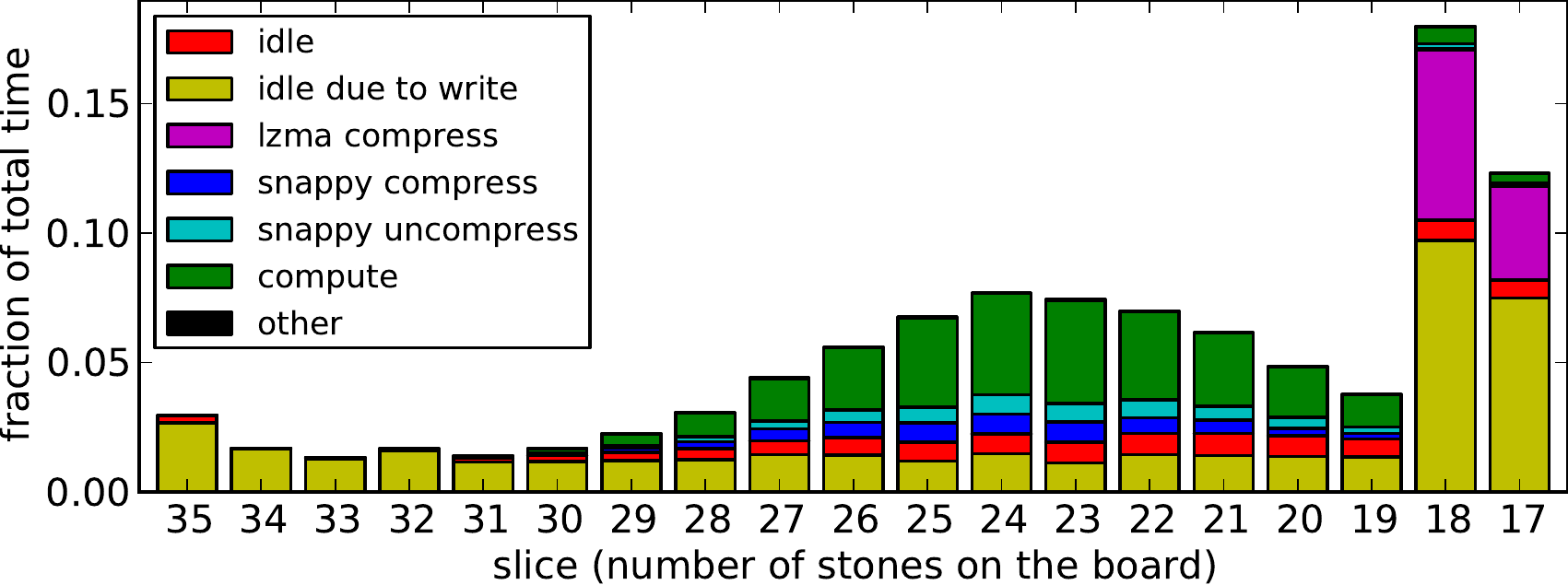}
\end{center}
\vspace{-.1in}
\cprotect\caption{Time profile of the main production run, showing total worker thread time usage over
all processes.  All game logic is contained in the \emph{compute} (green) section; the other sections are
overhead due to load imbalance, non-hidden communication latency, compression/decompression, and I/O.
Slices 17 and 18 include LZMA compression and final I/O.  During I/O, all workers are idle and the communication
thread on each process is inside MPI/IO.}
\label{profile}
\vspace{-.12in}
\end{figure}

The time profile of the computation is shown in \autoref{profile}.  Excluding I/O, only 16.4\% of the
total worker thread time is spent idle,\footref{idle}
confirming that our random load balancing scheme is sufficient for near peak performance.  Since only 5
out of 6 threads per rank are workers, we could theoretically speed up the computation by up to 6/5 if
the communication thread performed useful work.  Unfortunately, existing MPI implementations do not
implement performant asynchronous progress, even ignoring our need for active responses to messages
(see \cite{squyres2012progress} for a good discussion).

Including I/O, our performance is further from optimal: 51.0\% of worker thread time is
idle, with 34.6\% due entirely to I/O.  This is due to both very high latency when
writing small files during every slice (around 200 seconds independent of file size) and low bandwidth
when writing file results in slices 17 and 18.  The high latency was a consequence of
using \verb+MPI_File_write_ordered+ when writing small files, since MPICH and thus Cray MPI
implement this routine using shared files for synchronization rather than fast network collectives.
Unfortunately, the low bandwidth is likely user error: we accidentally wrote to NERSC's
global scratch filesystem rather than the special filesystem optimized for Edison.

Since much of the complexity of our implementation derives from the memory constrained in core
structure, it is important to estimate how much slower the computation would have been if run out of core.
Edison's peak I/O bandwidth is 168 GB/s, or 66 GB/s on our 2048 out of 5192 nodes if bandwidth is shared
proportionally.  An out of core version of our algorithm would write each block once and read each block
four times, for a total of 3.6 PB of I/O uncompressed or 0.94 PB with Snappy compression.  Thus, at peak I/O
bandwidth, a Snappy compressed out of core version of our code would take 4.0 hours for
I/O.\cprotect\footnote{See \verb+Total I/O time estimates+ in \verb+paper/numbers+.}  In contrast,
the non-I/O portion of our main run took 1.8 hours, for a speedup of 2.25.  If peak I/O performance could
not be achieved, or I/O and compute could not be fully overlapped, the speedup would be larger.

The rotation abstracted compute kernel uses SSE for instruction level parallelism, packing
each 256-bit $L \to \{0,1\}$ function into two 128 bit SSE registers.  $\rmax$ can be computed for one such
table in $152$ SSE instructions, or $3/5$ths of an instruction per position since each function
encodes $256$ positions.  Computing which of the 256 quadrant rotations of a board give five in
a row takes $190$ instructions and $640$ bytes of cache coherent table lookup, and transforming a
function $f_b$ into $f_{gb}$ takes between $60$ and $200$ instructions dependent on the particular $g \in G$.
Although bit-level representations of board state are standard in computational games,
we believe this is the first instance where values of many distinct positions are evaluated in parallel
using bit twiddling.

Measuring over the entire compute kernel (which excludes idle time, communication, and (de)compression),
our SSE routines achieve a $1.81\times$ speedup on Edison over 64-bit versions and a $2.25\times$ speedup on
an Intel Core i7, compared to a naive speedup of $2$ for twice as many bits per
instruction.\cprotect\footnote{See \verb+SSE vs. non-SSE speedup+ in \verb+paper/numbers+.}  We are not sure what
caused the superlinear speedup in the $2.25$ case; one possibility is reduced register pressure.

\begin{figure}
\begin{center}
\includegraphics[width=\columnwidth]{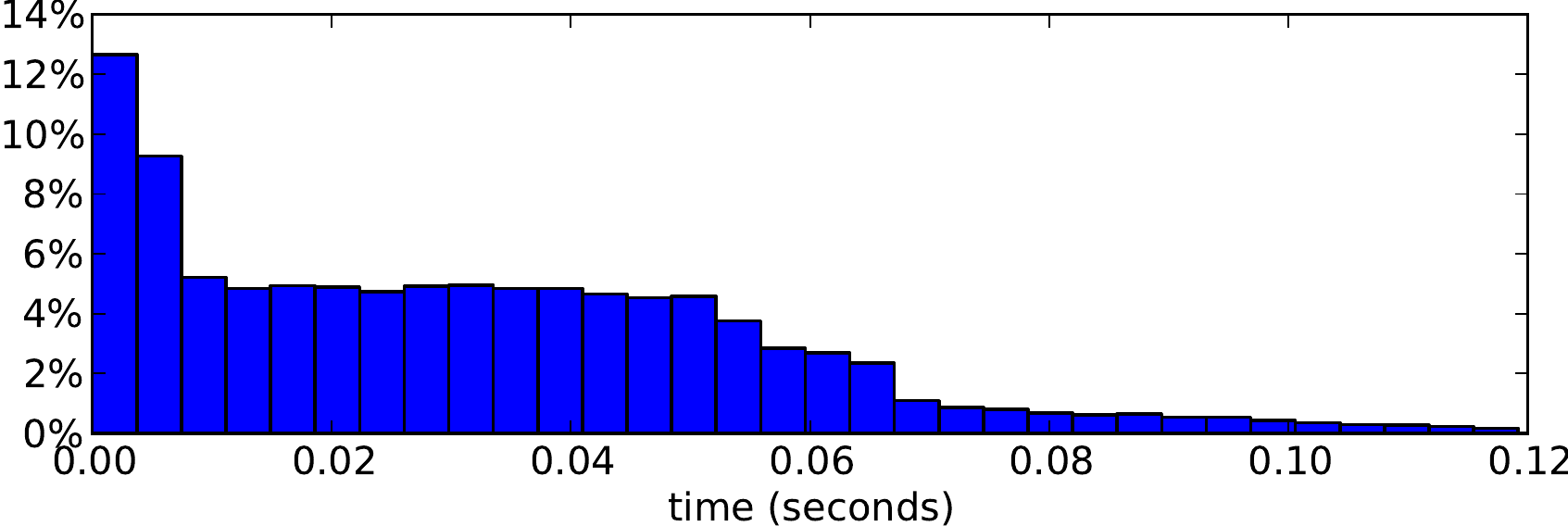}
\end{center}
\vspace{-.1in}
\cprotect\caption{Latency for 8-byte request-for-input messages between different nodes in a 96-thread test run
on Edison.  Around 20\% of the messages complete in under $10$ ms, but the tail is quite long.  To reproduce,
run \verb+paper/numbers messages 0,2+.}
\label{latency}
\vspace{-.1in}
\end{figure}

Even on Edison's faster network, our typical message latency is still quite high as shown in \autoref{latency}.
Here latency is measured from immediately before \verb+MPI_Isend+ to the time we start responding to the finished
message when \verb+MPI_Testsome+ succeeds.  The large latencies may be due to interference between small messages
and larger messages, as the communication thread processes different types of messages asynchronously.  Since
the latency on Edison is low enough for our purposes, we have not investigated in detail.

Within each node, we used the PAPI hardware counter library \cite{mucci1999papi} to measure instruction issue,
cache misses, and branch mispredications.  The results show that our workload has minimal memory
bandwidth requirements (under $15$ MB/s per core in performance critical sections), mispredicts branches mostly
during Snappy (de)compression, and makes significant use of dual instruction issue only for the one out of five
worker threads that share a core with the less active communication thread; if hyperthreading is turned off, the
issue rate jumps from $1.21$ to $1.89$ instructions per cycle.\cprotect\footnote{Run \verb+paper/numbers papi+.}

\section{Correctness and fault detection}

Since our goal is a database of perfect play, it is important to consider the possible sources of error in the
computation.  We are interested only in undetected errors, since empirically the code ran
without crashing or failing an assertion.

On the software side, we make heavy use of unit tests throughout the code, including simple tests
for correctness of simple routines, bootstrap tests comparing simple routines to more complicated variants
(such as when abstracting over rotations), and comparison tests between different algorithms.
In particular, we compare our parallel backward code against the results of forward search, both at the
beginning and end of the game.  Near the end of the game this is easy, as forward search can quickly compute
perfect play.  Near the beginning, our tests replace all values at slice 4 or 5 with random values,
compute optimal play up to the random slice with both backward and forward algorithms, and compare.

On the hardware side, the main failure points are DRAM, CPU, network, and disk.  Undetected disk errors are unlikely
since the primary output files are checksummed as part of LZMA compression.  Of DRAM, CPU, and network, DRAM errors
dominate according to \cite{bergman2008exascale} (Tables 6.11 and 6.12), at least on BlueGene.  Edison memory
and network are SECDED (single error correct, double error detect), so an undetectable error would require three
simultaneous failures.  Unfortunately, we do not know a reliable method to estimate this probability conditional
on an apparently successful run, since DRAM errors are far from uncorrelated events \cite{schroeder2009dram}.
However, we believe the probability is quite small, and in particular that undetected hardware errors are less
likely than software errors.

As a test on software errors when running the code at scale, we write out small sample files with the
results of randomly chosen boards during each slice.  Large numbers of samples generated by the main run were
validated against forward search for slices 20 and up, so any remaining software errors in the parallel code
must manifest only on a small set of positions.  We also write out win/loss/tie counts for each slice.
Both sample and count files would be useful for cross-validation should someone reproduce the calculation
in the future.

Unfortunately, the sample files are insufficient to detect rare software bugs or hardware failures, and indeed
we know of no cheap method for detecting this kind of unlikely error without rerunning or rewriting the code.
Solving pentago falls most naturally into the complexity class PSPACE (polynomial space), and indeed the similar
five-in-a-row game gomoku has been proven PSPACE-complete \cite{reisch1980gobang}.  Unless
$\textrm{NP}=\textrm{PSPACE}$, it is unlikely that a short certificate exists proving that pentago is
a first player win, especially if we require a strong solution with perfect play known from all positions.

\section{Open source and data}

All code for this project is open source, and is available online at \url{https://github.com/girving/pentago}.
The repository includes the paper source and all log files used to generate timing and other numbers.
To regenerate any reported number from the data, run either
\verb+bin/analyze <command>+ or \verb+paper/numbers+; see the footnotes and figure captions.

The 3.7 TB strong solution is hosted on Rackspace Cloud Files; see the download instructions at
\url{https://github.com/girving/pentago/#data}.  We store small sparse sample and count files in
Numpy's \verb+.npy+ format \cite{kern2007npy}, and the main solution files in a custom \verb+.pentago+ format
using the described block structure with LZMA compression per block.  The format is described
at \url{https://github.com/girving/pentago/blob/master/pentago/data/supertensor.h}.  Both \verb+.npy+ and
\verb+.pentago+ formats are easy to write in parallel using MPI I/O.

The strong solution is useless without a convenient method for exploring the data, so we have built a website
showing which moves win, lose, or tie from any position: \url{http://perfect-pentago.net}.  The frontend
Javascript uses a backend server at \url{http://backend.perfect-pentago.net:2048} to look up the value of
positions.  Any position with 18 or fewer stones is fetched from the database using an HTTP range request
to download the surrounding compressed block.  As in the parallel algorithm, the children of a position
fall into at most four blocks; we cache the uncompressed blocks to take advantage of this locality.

Positions with more than 18 stones fall outside the database and are recomputed from scratch using a specialized
serial retrograde solver.  Since there are at least 18 stones already on the board, usually in an asymmetric
configuration, this solver rotates the board only through the $\rmax$ function, avoiding the complexity of
standardizing positions into rotation minimal configurations.  In addition, we use the fact that $\rmax$ flips
the parity of the $\Z_4^4$ symmetry group to store half the required bits, reducing the storage per rotation
abstracted position from 64 bytes to 32.  With these optimizations, evaluating all child values of an 18
stone position takes 16 seconds on a single 2.6 GHz Intel Xeon thread, fast enough for interactive use.

Both remote lookups and from-scratch computation have significant latency, so the backend server is written
in Javascript using Node.js \cite{dahl2014nodejs} for asynchronous use by multiple clients.  The Javascript
handles asynchronous logic and I/O, but calls down to C++ for performance intensive computation.  The backend
server has a simple JSON API, and anyone wishing to develop their own frontend is welcome to query it directly.

\section{Conclusion}

We have strongly solved the board game pentago using retrograde analysis on 98304 threads of Edison, a
Cray XC30 machine at NERSC.  Symmetry techniques were used to improve the branching factor of the game
and take advantage of SSE instruction level parallelism.  Unlike previous retrograde analysis, the computation
was almost entirely in-core, writing results to disk only near the end.  To fit safely into
memory, we use a fully asynchronous communication structure where each process requests data from other
processes as needed, performs computation, and scatters results to their destinations.

The asynchronous control flow was a primary complicating factor during development and optimization of the
code, and runs against several limitations of MPI including difficulties in synchronizing between MPI and threads,
lack of support for asynchronous progress in existing implementations, poor one-sided communication in MPI 2
(fixed in MPI 3 too late for use in this project), and lack of user defined operations in one-sided
\verb+MPI_Accumulate+.  Unfortunately, the latter would require both strong asynchronous progress and careful
consideration of threading semantics.  Profiling tools were also a significant limitation, leading us to implement
our own tracing and visualization tool to understand the flow of information across processes and between threads
without one process.  Although our custom tool helped localize the problem to high latency, we were unable to
diagnose the underlying cause; further analysis would likely require network profiling and visualization tools
incorporating knowledge of network topology.

Compression was a requirement to fit into memory, but we were limited to the fast and weak Snappy library to
prevent compression from becoming a compute bottleneck.  Compression also makes memory usage difficult to
predict in advance, causing us to overestimate memory requirements and use more Edison nodes than required.
Avoiding such overestimate without the I/O cost of checkpointing would require a dynamic number of MPI nodes.

Our computation shares many characteristics with other irregularly structured, data intensive HPC workloads.
These characteristics include multiple levels of structure (slices, sections, block lines, blocks, boards,
bits), memory restrictions, asynchronous control flow, reliance on integer performance (compression and game
logic), and reliance on both fast compute and fast communication.  Multiple levels of structure are important
in many applications (e.g., domains, pages, paragraphs, sentences, words for web search) and often warrant
different parallelism strategies at different levels.  In addition to allowing larger problem sizes either
in RAM or on-package RAM, the ability to operate near a memory limit improves performance for codes with
imperfect parallel scaling and eases the scheduling problem for shared clusters, increasing both latency and
bandwidth for users.  Asynchronous control flow adds flexibility which can be spent on memory constraints or
irregular work chunk sizes (common with multiple levels of structure).  Poor integer performance ruled out
BlueGene for our purposes, which is problematic even for floating point codes if compression is required.
Finally, traditional Big Data applications often have less tightly coupled communication patterns such as
MapReduce \cite{dean2008mapreduce}; our application is sufficiently latency-critical to obtain clear benefit
from the faster network on Edison compared to Hopper, and serves as an intermediate example between traditional
HPC and Big Data (see the Graph 500 benchmark suite for other examples \cite{murphy2010graph500}).

\section*{Acknowledgments}

I am grateful to Jeff Hammond for valuable advice throughout the project, and to Jed Brown for numerous helpful
suggestions including the initial suggestion that supercomputer time might be a possibility.
Hosting for the 3.7 TB data set and compute servers for \url{http://perfect-pentago.net} were generously donated
by Rackspace; thanks especially to Jesse Noller at Rackspace for offering to host this open source project.
Since a primary goal of this project is open data, accessible hosting for the final results is essential.
This research used resources of the National Energy Research Scientific Computing Center, which is supported by
the Office of Science of the U.S. Department of Energy under Contract No.\ DE-AC02-05CH11231.  This research also
used resources of the Argonne Leadership Computing Facility at Argonne National Laboratory, which is supported by
the Office of Science of the U.S. Department of Energy under contract DE-AC02-06CH11357.

\bibliography{references}
\bibliographystyle{IEEEtran}
\end{document}